\documentclass[10pt,aps,twocolumn,superscriptaddress,floatfix,prc,twoside,tightenlines,nofootinbib,showpacs]{revtex4}
\usepackage[sort&compress]{natbib}
\usepackage{graphicx,mathrsfs,amsmath,amssymb,amsopn,mathbbol,bm,qmech-revtex}
\usepackage{color}

\newcommand{\ie}{\textit{i.e.}}
\begin{document}

\title{Thermal Dileptons as Fireball Thermometer and Chronometer} 
\author{Ralf Rapp} 
\affiliation{Cyclotron Institute and Physics Department, Texas A{\&}M
  University, College Station, Texas 77843-3366, USA} 
\author{Hendrik van Hees}
\affiliation{Institut f{\"u}r Theoretische Physik,
  Johann-Wolfgang-Goethe-Universit{\"a}t Frankfurt, Max-von-Laue-Str.\ 1,
  D-60438 Frankfurt, Germany}
\affiliation{Frankfurt Institute for Advanced Studies,
  Ruth-Moufang-Str.\ 1,  D-60438 Frankfurt, Germany}

\date{\today}

\begin{abstract}
  Thermal dilepton radiation from the hot fireballs created in
  high-energy heavy-ion collisions provides unique insights into the
  properties of the produced medium. We first show how the predictions
  of hadronic many-body theory for a melting $\rho$ meson, coupled with
  QGP emission utilizing a modern lattice-QCD based equation of state,
  yield a quantitative description of dilepton spectra in heavy-ion
  collisions at the SPS and the RHIC beam energy scan program. We
  utilize these results to systematically extract the excess yields and
  their invariant-mass spectral slopes to predict the excitation
  function of fireball lifetimes and (early) temperatures,
  respectively. We thereby demonstrate that future measurements of these
  quantities can yield unprecedented information on basic fireball
  properties. Specifically, our predictions quantify the relation
  between the measured and maximal fireball temperatures, and the
  proportionality of excess yields and total lifetime. This information
  can serve as a ``caloric'' curve to search for a first-order QCD phase
  transition, and to detect non-monotonous lifetime variations possibly
  related to critical phenomena.
\end{abstract}

\maketitle


Collisions of heavy nuclei at high energies enable the creation of hot
and dense strongly interacting matter, not unlike the one that filled
the Universe during its first few microseconds. While the primordial
medium was characterized by a nearly vanishing net baryon density (at
baryon chemical potential $\mu_B$$\simeq$0), heavy-ion collisions can
effectively vary the chemical potential by changing the beam energies,
thus facilitating systematic investigations of large parts of the phase
diagram of Quantum Chromodynamics (QCD). The yields and
transverse-momentum ($p_T$) spectra of produced hadrons have been widely
used to determine the conditions of the fireball at chemical and kinetic
freezeout, to infer its properties when the hadrons decouple.
Electromagnetic radiation (photons and dileptons), on the other hand, is
emitted throughout the evolution of the expanding fireball with
negligible final-state interactions and thus, in principle, probes the
earlier hotter phases of the medium~\cite{Shuryak:1978ij}. In
particular, dilepton invariant-mass spectra have long been recognized as
the only observable which gives direct access to an in-medium spectral
function of the QCD medium, most notably of the $\rho$
meson~\cite{Brown:1991kk,Pisarski:1995xu,Rapp:1999ej,Harada:2003jx}.
They also allow for a temperature measurement which is neither distorted
by blue-shift effects due to collective expansion (as is the case for
$p_T$ spectra of hadrons and photons), nor limited by the hadron
formation temperature~\cite{BraunMunzinger:2003zz}.

Significant excess radiation of dileptons in ultrarelativistic heavy-ion
collisions (URHICs), beyond final-state hadron decays, was established
at the CERN-SPS program, at collision energies of
$\sqrt{s_{NN}}\simeq20\,\GeV$~\cite{Tserruya:2009zt,Specht:2010xu}. The
excess was found to be consistent with thermal radiation from a locally
equilibrated fireball~\cite{Rapp:1999us,Rapp:1999zw}, with the low-mass 
spectra requiring substantial medium effects on the $\rho$ line shape. 
The SPS dilepton program culminated in the high-precision NA60 data, which 
quantitatively confirmed the melting of the $\rho$ resonance and realized 
the long-sought thermometer at masses $M> 1\,\GeV$, with $T$=205$\pm$12\,MeV, 
exceeding the pseudo-critical temperature computed in thermal lattice-QCD 
(lQCD), $T_{\mathrm{pc}}$=150-170\,MeV~\cite{Borsanyi:2010bp}. With the
spectral shape under control, the magnitude of low-mass excess enabled an 
unprecedented extraction of the fireball lifetime, 
$\tau_{\rm FB}$=7$\pm$1\,fm/$c$.

In the present letter, based on a good description of existing dilepton
data from CERES, NA60 and STAR, we show that temperature and lifetime
measurements through intermediate- and low-mass dileptons are a
quantitative tool to characterize the fireballs formed in heavy-ion
collisions. We predict pertinent excitation functions over a large range
in center-of-mass energy, $\sqrt{s_{NN}}$$\simeq$6-200\,GeV. The 
motivation for this study is strengthened by several recent developments. 
First, we show that the implementation of a modern lQCD equation of state 
(EoS) into the fireball evolution of In-In collisions at SPS recovers an 
accurate description of the NA60 excess data over the entire mass range, 
thus superseding earlier results with a first-order 
transition~\cite{vanHees:2007th}. Second, the predictions of this framework 
turn out to agree well with the low-mass dilepton data from the STAR 
beam-energy scan-I (BES-I) campaign~\cite{Geurts:2012rv,Huck:2014mfa}, 
in Au-Au collisions from SPS to top RHIC energies, 
$\sqrt{s_{NN}}=19.6$, 27, 39, 62.4 and $200\, \GeV$~\cite{Rapp:2013nxa}.  
Third, a very recent implementation of the in-medium $\rho$ spectral 
function into coarse-grained UrQMD transport calculations yields excellent 
agreement with both NA60 and HADES data in Ar-KCl($\sqrt{s_{NN}}=2.6\,\GeV$)
reactions~\cite{Endres:2014}; and fourth, several future experiments
(CBM, NA60+, NICA, STAR BES-II) plan precision measurements of dilepton
spectra in the energy regime of 
$\sqrt{s_{NN}}\simeq 5$-$20\,\GeV$~\cite{Friman:2011zz}, where the fireball 
medium is expected to reach maximal baryon density and possibly come close
to a critical point in the QCD phase diagram~\cite{Stephanov:2007fk}. Our 
predictions thus provide 
a baseline for fundamental, but hitherto undetermined properties of the 
fireball, allowing for accurate tests of our understanding of these. In 
turn, marked deviations of upcoming data from the theoretical predictions 
will help discover new phenomena that induce unexpected structures in the 
lifetime and/or temperature excitation functions.
 
Let us recall the basic elements figuring into our calculation of dilepton
excess spectra in URHICs. We assume local thermal equilibrium of the fluid 
elements in an expanding fireball, after a (short) initial equilibration 
period. The thermal radiation of dileptons is obtained from the differential 
production rate per unit four-volume and four-momentum~\cite{MT84,wel90,gale-kap90},
\begin{equation}
\label{del-rate}
\frac{\dd N_{ll}}{\dd^4 x \dd^4 q}
=-\frac{\alpha^2}{3 \pi^3} \frac{L(M)}{M^2}~\im
\Pi_{{\rm EM},\mu}^{\mu}(M,q)~f_{\mathrm{B}}(q_0;T),
\end{equation}
with $f_{\mathrm{B}}(q_0;T)$: thermal Bose function,
$\alpha$=$\frac{e^2}{4 \pi}$$\simeq$$\frac{1}{137}$: electromagnetic
(EM) coupling constant, $L(M)$: final-state lepton phase space factor,
and $M$=$\sqrt{q_0^2-q^2}$: dilepton invariant mass ($q_0$: energy, $q$:
three-momentum in the local rest frame of the medium).
The EM spectral function, Im\,$\Pi_{\rm EM}$, is well known in the
vacuum, being proportional to the cross section for $\ee^+\ee^-\to \
\text{hadrons}$. In the low-mass region (LMR), $M\le 1 \, \GeV$, it is
saturated by the light vector mesons $\rho$, $\omega$ and $\phi$,
while the intermediate-mass region (IMR), $M \ge 1.5\,\GeV$, is
characterized by a continuum of multi-meson states.

Medium effects on the EM spectral function are calculated as follows. In
hadronic matter the vector-meson propagators are computed using
many-body theory based on effective Lagrangians with parameters
constrained by vacuum scattering and resonance decay
data~\cite{Urban:1998eg,Rapp:1999us,Rapp:2000pe}.  The resulting $\rho$
spectral function strongly broadens and melts in the phase transition
region (similar for the $\omega$, which, however, contributes less than
10\%; the $\phi$ is assumed to decouple near $T_{\rm pc}$ and does not
produce thermal hadronic emission). For masses $M \ge 1\,\GeV$, we
include emission due to multi-pion annihilation using a continuum
extracted from vacuum $\tau$-decay data, augmented with medium effects
due to chiral mixing~\cite{Dey:1990ba,SYZ96} in the LMR-IMR transition
region (1\,GeV~$\le M \le$~1.5\,GeV)~\cite{vanHees:2007th}.  For QGP
emission, we employ an lQCD-motivated emission
rate~\cite{Rapp:2013nxa} fitted to $M$-dependent spectral functions
above $T_{\rm pc}$~\cite{Ding:2010ga,Brandt:2012jc} and supplemented by
a finite-$q$ dependence taken from perturbative photon
rates~\cite{Kapusta:1991qp}. The resulting QGP rates are quite similar
to the hard-thermal-loop results~\cite{Braaten:1990wp}, but with
improved low-mass behavior and nontrivial three-momentum dependence. The
QGP and in-medium hadronic rates are nearly degenerate at temperatures
around $\sim$170\,MeV.

\begin{table}[!t]
\begin{center}
\begin{tabular}{|c||c|c|c|c|c|}
\hline
$\sqrt{s}$ (GeV)  &  6.3 & 8.8  & 19.6 & 62.4 & 200 
\\
\hline  \hline
$z_0$ (fm/$c$) & 2.1 & 1.87 & 1.41 & 0.94 & 0.63
\\
\hline
$T_{\rm pc}$ (MeV) & 161  & 163 & 170 & 170 & 170 
\\
\hline
$T_{\rm ch}$ (MeV) & 134  & 148& 160 & 160 & 160 
\\
\hline
$\mu_B^{\rm ch}$ (MeV)  & 460 & 390 & 197  & 62 & 22 
\\
\hline
$T_{\rm kin}$ (MeV) & 113  & 113 & 111 & 108 & 104 
\\
\hline
\end{tabular}
\end{center}
\caption{Excitation function of fireball parameters for the equation of
state ($T_{\rm pc}$), initial ($z_0$) and chemical/kinetic 
freezeout ($T_{\rm ch}$, $\mu_B^{\rm ch}$, $T_{\rm kin}$) conditions.}
\label{tab-eos}
\end{table}
\begin{figure}[!t]
\centerline{\includegraphics[width=0.95\columnwidth]{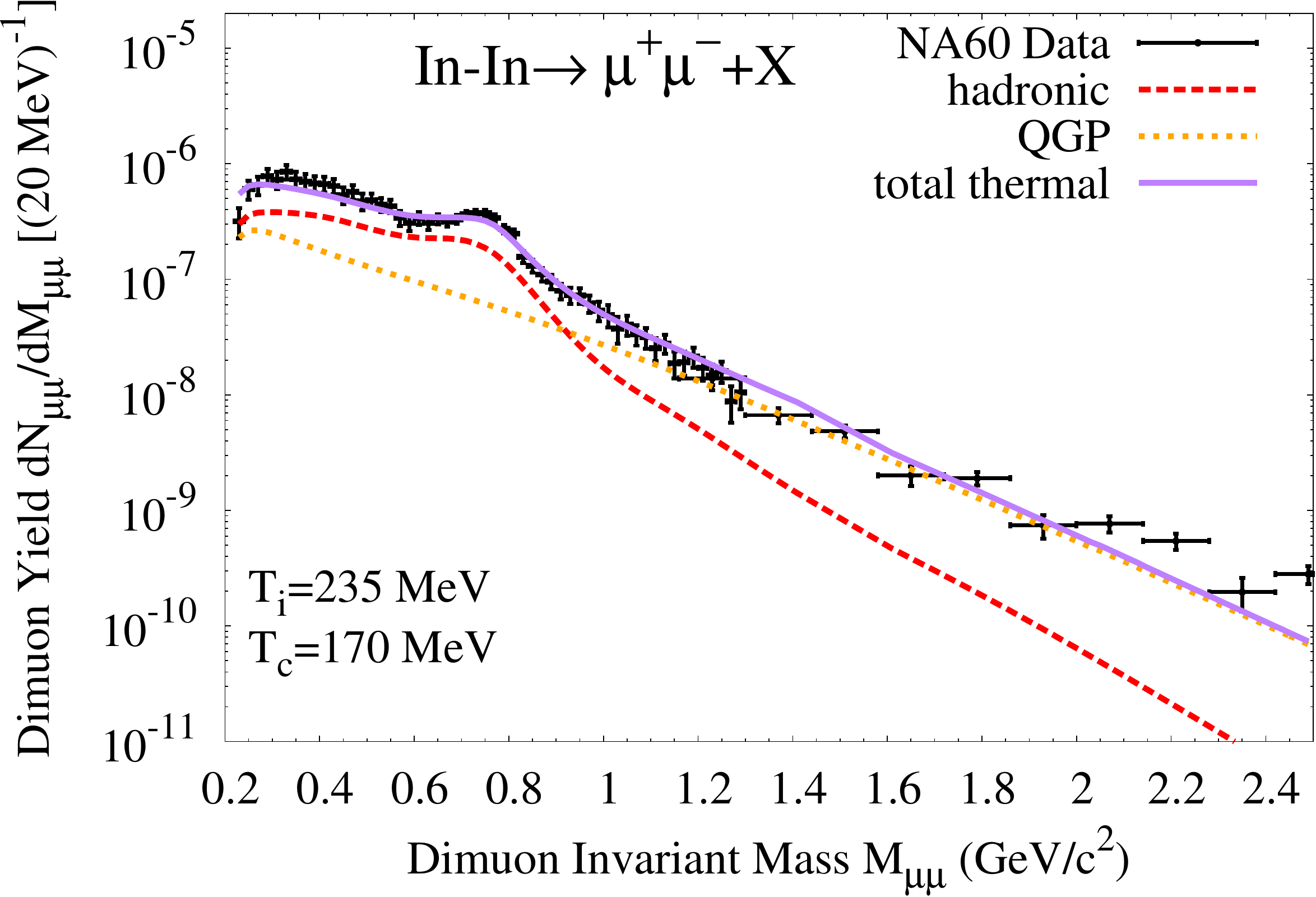}}
\caption{(Color online) Excess dimuon invariant-mass spectra in
  In-In($\sqrt{s_{NN}}=17.3\,\GeV$) collisions at the
  SPS. Theoretical calculations (solid line), composed of hadronic
  radiation (using in-medium $\rho$ and $\omega$ spectral functions and
  multi-pion annihilation with chiral mixing, dashed line) and QGP
  radiation (using a lattice-QCD inspired rate, dotted line) are
  compared to NA60 data~\cite{Arnaldi:2008fw,Specht:2010xu}.}
\label{fig_na60}
\end{figure}

To obtain dilepton spectra in URHICs, the rates are integrated
over the space-time evolution of the collision. As in our previous
work~\cite{vanHees:2007th,vanHees:2011vb,Rapp:2013nxa,vanHees:2014ida},
we employ a simplified model in terms of an isentropically expanding
thermal fireball. Its radial and elliptic flow are parameterized akin to
hydrodynamic models and fitted to observed bulk-particle spectra and
elliptic flow ($\pi$, $K$, $p$) at kinetic freezeout,
$T_{\mathrm{kin}}$$\simeq$100-120\,MeV, and to multistrange hadron
observables (e.g., $\phi$) at chemical freezeout, $T_{\rm
  ch}$$\simeq$160\,MeV.  The kinetic freezeout temperatures and radial
flow velocities are in accordance with systematic blast-wave analyses 
of bulk-hadron spectra from SPS, RHIC and LHC~\cite{Kumar:2014tca}.
The key link of the fireball expansion to dilepton emission is the
underlying EoS, which converts the time-dependent entropy density,
$s(\tau)=S/V_{\mathrm{FB}}(\tau)\equiv s(T(\tau),\mu_B(\tau))$
($V_{\mathrm{FB}}$: fireball volume), into temperature and chemical
potential. We employ the EoS constructed in Ref.~\cite{He:2011zx}, where
a parameterization of the $\mu_B$=0 lQCD results for the
QGP~\cite{Borsanyi:2010cj,Cheng:2009zi} has been matched to a hadron
resonance gas (HRG) at $T_{\mathrm{pc}}$=170\,MeV, with subsequent
hadro-chemical freezeout at $T_{\mathrm{ch}}$=160\,MeV.  We here extend
this construction to finite $\mu_B$=3$\mu_q$ with guidance from lQCD:
The pseudo-critical temperature is reduced as $T_{\rm
  pc}(\mu_q)$=$T_{\rm pc} [1 - 0.08 (\mu_q/T_{\rm pc})^2]$
\cite{Endrodi:2011gv,Kaczmarek:2011zz}, and the QGP EoS is modified as
$s(\mu_q,T)=s(T) [1+c (\mu_q/\pi T)^2]$ with
$c$$\simeq$3~\cite{Borsanyi:2012cr,Hegde:2014sta}.  For the HRG, we
adopt the chemical freezeout parameters of
Ref.~\cite{BraunMunzinger:2011ze}, cf.~Tab.~\ref{tab-eos}.

We first test our updated approach with the most precise dilepton data
available, the acceptance-corrected NA60 excess dimuons in
In-In($\sqrt{s_{NN}}$=17.3\,GeV)~\cite{Arnaldi:2008fw,Specht:2010xu},
cf.~Fig.~\ref{fig_na60}. Good agreement with the invariant-mass spectrum
is found, which also holds for the $q_{\mathrm{t}}$ dependence, as well
as for CERES data~\cite{Agakichiev:2005ai} (not shown). This confirms
our earlier conclusions that the $\rho$-meson melts around $T_{\rm
  pc}$~\cite{vanHees:2007th}, while the IMR is dominated by radiation from 
above $T_{\rm pc}$~\cite{Renk:2006qr,Dusling:2006yv,Alam:2009da,Linnyk:2011hz},
mostly as a consequence of a nonperturbative EoS~\cite{Rapp:2013ema}. Furthermore, 
our predictions for low-mass
and $q_t$ spectra of the RHIC BES-I program~\cite{Rapp:2013nxa} agree
well with STAR dielectron data~\cite{Geurts:2012rv,Huck:2014mfa}. Given
this robust framework for thermal dilepton radiation in URHICs,
we extract in the following the excitation function of two key fireball
properties, namely its total lifetime and an average temperature, 
directly from dilepton observables.

\begin{figure}[!t]
\vspace{-0.8cm}
\centerline{\includegraphics[width=1.0 \columnwidth]{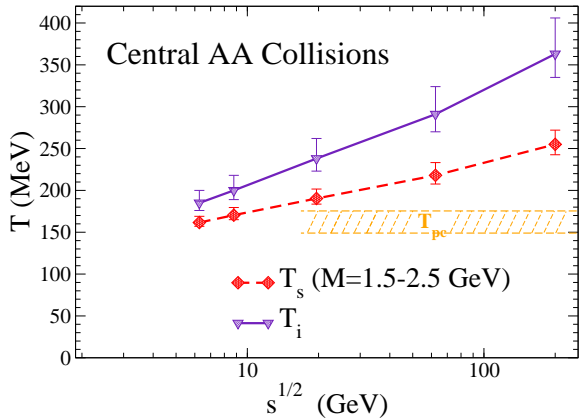}}
\caption{(Color online) Excitation function of the inverse-slope
  parameter, $T_{\rm s}$, from intermediate-mass dilepton spectra
  ($M$=1.5-2.5\,\GeV, diamonds connected with dashed line) and initial
  temperature (triangles connected with solid line) in central heavy-ion
  collisions ($A$\,$\simeq$\,200). The hatched area schematically indicates
  the pseudo-critical temperature regime at vanishing (and small)
  chemical potential as extracted from various quantities computed in
  lattice QCD~\cite{Borsanyi:2010bp}.}
\label{fig_slopes}
\end{figure}
For the temperature determination we utilize the IMR, where medium
effects on the EM spectral function are parametrically small, of order
$T^2/M^2$, providing a stable thermometer: With $\im \Pi_{\rm EM}\propto
M^2$, and in nonrelativistic approximation, one obtains $\dd R_{ll}/\dd
M\propto (MT)^{3/2}\exp(-M/T)$, which is {\em independent} of the
medium's collective flow, \ie, there are no blue-shift effects. The
observed spectra necessarily involve an average over the fireball
evolution, but the choice of mass window, $1.5 \, \GeV \le M \le 2.5 \,
\GeV$, implies $T\ll M$ and thus enhances the sensitivity to the early
high-$T$ phases of the evolution. Since primordial (and pre-equilibrium)
contributions are not expected to be of exponential shape (e.g., power
law for Drell-Yan), their ``contamination'' may be judged by the
fit quality of the exponential ansatz. The inverse slopes, $T_{\rm s}$,
extracted from the thermal radiation as computed above are displayed in
Fig.~\ref{fig_slopes} for collision energies of
$\sqrt{s_{NN}}$=6-200\,GeV.  We find a smooth dependence ranging from
$T$$\simeq$160\,MeV to 260\,MeV.  The latter unambiguously demonstrates
that a thermalized QGP with temperatures well above the pseudo-critical
one has been produced. Our results furthermore quantify that the
``measured'' average temperature is about 30\% below the corresponding
initial one ($T_i$). This gap significantly decreases when lowering the
collision energy, to less than 15\% at $\sqrt{s_{NN}}$=6\,GeV. This is
in large part a consequence of the (pseudo-) latent heat in the
transition which needs to be burned off in the expansion/cooling. The
collision energy range below $\sqrt{s_{NN}}$=10\,GeV thus appears to be
well suited to map out this transition regime and possibly discover a 
plateau in the IMR dilepton slopes akin to a ``caloric curve". Another 
benefit at these energies is the smallness of the open-charm contribution
(not included here), so that its subtraction does not create a large
systematic error in the thermal-slope measurement. The main theoretical
uncertainty in our calculations is associated with the assumed initial
longitudinal fireball size, $z_0$ (which is proportional to the
thermalization time): varying the default values quoted in
Table~\ref{tab-eos} by $\pm$30\% induces a $\sim$5-7\% change in the
extracted slopes, and a somewhat larger change for the initial temperatures.  
At given $\sqrt{s}$, the ratio $T_{\rm s}/T_{\rm i}$ is stable within less
than 10\%.

\begin{figure}[!t]
\centerline{\includegraphics[width=0.95\columnwidth]{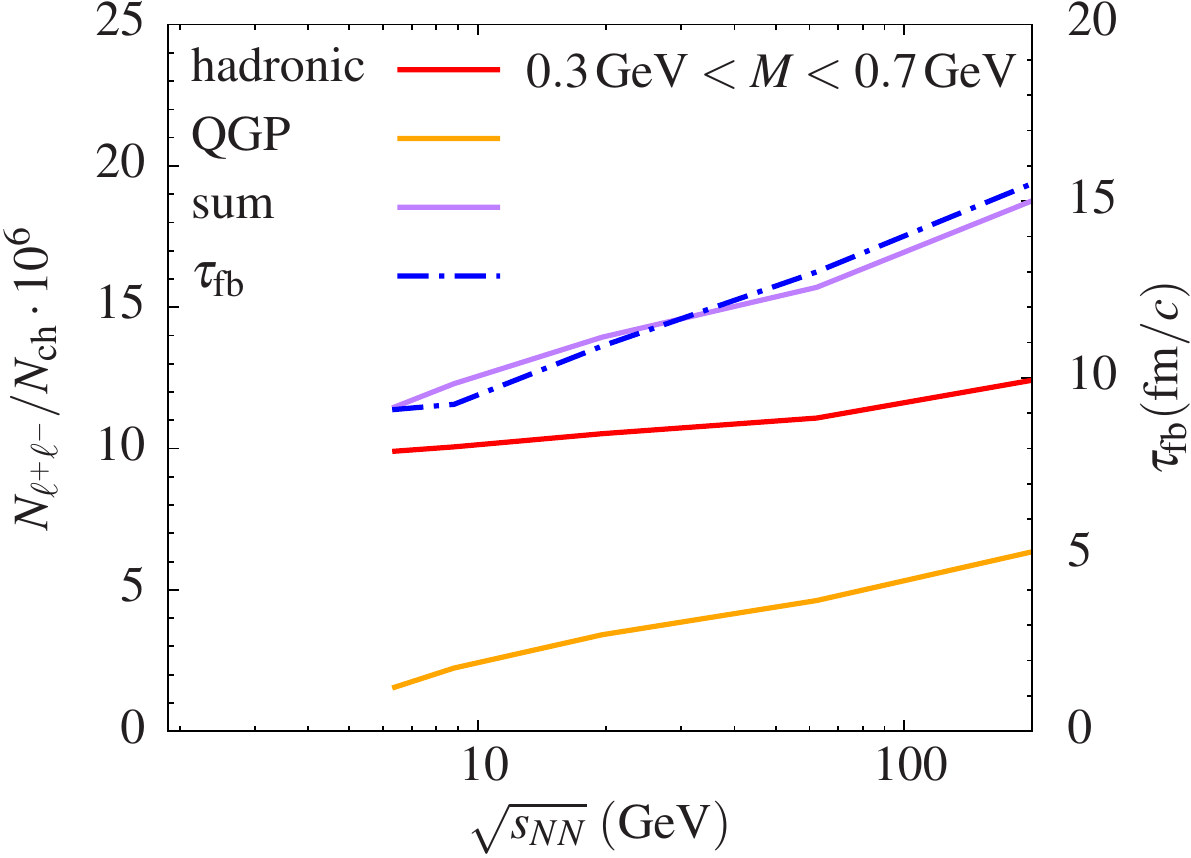}}
\caption{(Color online) Excitation function of low-mass thermal
  radiation (``excess spectra'') integrated over the mass range
  $M$=0.3-0.7\,GeV, as given by QGP (orange line) and in-medium hadronic
  (red line) contributions and their sum (purple line). The underlying
  fireball lifetime (dot-dashed line) is given by the right vertical
  scale.}
\label{fig_excess}
\end{figure}
We finally investigate the relation between the fireball lifetime and
the thermal dilepton yields, integrated over a suitable mass window. In
Fig.~\ref{fig_excess} we display the results for a window below the free
$\rho$/$\omega$ mass, which is often used to characterize the low-mass
excess radiation. It turns out that the integrated thermal excess
radiation tracks the total fireball lifetime remarkably well, within less
than 10\%.  An important reason for this is that, despite the dominantly
hadronic contribution, the QGP one is still significant. The latter
would be relatively more suppressed when including the $\rho$/$\omega$
peak region. Likewise, the hadronic medium effects are essential to
provide sufficient yield in the low-mass region. We have explicitely
checked that when hadronic medium effects are neglected, or when the
mass window is extended, the proportionality of the excess yield to the
lifetime is compromised. With such an accuracy, low-mass dileptons are
an excellent tool to detect any ``anomalous'' variations in the
fireball lifetime. A good control over the in-medium spectral
shape is essential here, at the level established in the comparison to
the NA60 data in Fig.~\ref{fig_na60}.

In summary, we have computed thermal dilepton spectra in heavy-ion
collisions over a wide range of collision energies, utilizing in-medium
QGP and hadronic emission rates in connection with a lattice-QCD
equation of state extrapolated to finite chemical potential. Our
description satisfies the benchmark of the high-precision NA60 data at
the SPS and is compatible with the recent results from the RHIC
beam-energy scan. Within this framework, we have extracted the
excitation function of the low-mass excess radiation and the
Lorentz-invariant slope of intermediate-mass spectra. The former turns
out to accurately reflect the average fireball lifetime. The latter
signals QGP radiation well above the critical one at top RHIC energy,
but closely probes the transition region for center-of-mass energies
below 10\,GeV. Dilepton radiation is thus well suited to provide direct
information on the QCD phase boundary in a region where a critical point
and an onset of a first-order transition are conjectured.

\begin{acknowledgments}
This work was supported in part by a U.S. National Science Foundation 
under grant PHY-1306359, by the Humboldt foundation (Germany), by
BMBF and LOEWE.
\end{acknowledgments}


\begin{thebibliography}{99}

\bibitem{Shuryak:1978ij} 
  E.V.~Shuryak,
  Phys.\ Lett.\ B {\bf 78}, 150 (1978).

\bibitem{Brown:1991kk} 
  G.E.~Brown and M.~Rho,
  Phys.\ Rev.\ Lett.\  {\bf 66}, 2720 (1991).

\bibitem{Pisarski:1995xu} 
  R.D.~Pisarski,
  Phys.\ Rev.\ D {\bf 52}, 3773 (1995).

\bibitem{Rapp:1999ej}
  R.~Rapp and J.~Wambach,
  Adv.\ Nucl.\ Phys.\  {\bf 25}, 1 (2000).

\bibitem{Harada:2003jx} 
  M.~Harada and K.~Yamawaki,
  Phys.\ Rept.\  {\bf 381}, 1 (2003).

\bibitem{BraunMunzinger:2003zz} 
  P.~Braun-Munzinger, J.~Stachel and C.~Wetterich,
  Phys.\ Lett.\ B {\bf 596}, 61 (2004).

\bibitem{Tserruya:2009zt}
  I.~Tserruya,
in {\em Relativistic Heavy-Ion Physics}, edited by R.~Stock and
  Landolt B\"ornstein (Springer), New Series {\bf I/23A} (2010) 4-2
  [arXiv:0903.0415[nucl-ex]].

\bibitem{Specht:2010xu}
  H.J.~Specht  [NA60 Collaboration],
  AIP Conf.\ Proc.\  {\bf 1322}, 1 (2010).

\bibitem{Rapp:1999us}
  R.~Rapp and J.~Wambach,
  Eur.\ Phys.\ J.\ A {\bf 6}, 415 (1999).

\bibitem{Rapp:1999zw} 
  R.~Rapp and E.~V.~Shuryak,
  Phys.\ Lett.\ B {\bf 473}, 13 (2000).

\bibitem{Borsanyi:2010bp}
  S.~Borsanyi {\it et al.},  
  JHEP {\bf 1009}, 073 (2010).

\bibitem{vanHees:2007th}
H.~van Hees and R.~Rapp,
Nucl.\ Phys.\ A  \textbf{806}, 339 (2008).

\bibitem{Geurts:2012rv}
  F.~Geurts [STAR Collaboration],
  Nucl.\ Phys.\ A {\bf 904-905}, 217c (2013).

\bibitem{Huck:2014mfa} 
  P.~Huck [STAR Collaboration],
  arXiv:1409.5675[nucl-ex].

\bibitem{Rapp:2013nxa} 
  R.~Rapp,
  Adv.\ High Energy Phys.\  {\bf 2013}, 148253 (2013).

\bibitem{Endres:2014}
S.~Endres, H.~van Hees and M.~Bleicher, in prep. (2014). 

\bibitem{Friman:2011zz} 
  B.~Friman {\it et al.}, 
  Lect.\ Notes Phys.\  {\bf 814}, 1 (2011).

\bibitem{Stephanov:2007fk} 
  M.~A.~Stephanov,
  PoS LAT {\bf 2006}, 024 (2006).

\bibitem{MT84}
L.~D. McLerran and T.~Toimela, Phys. Rev. D \textbf{31}, 545 (1985).
                                                                                          
\bibitem{wel90}
H.~A. Weldon, Phys. Rev. D \textbf{42}, 2384 (1990).
                                                                                          
\bibitem{gale-kap90}
C.~Gale and J.~I. Kapusta, Nucl. Phys. B \textbf{357}, 65 (1991).

\bibitem{Urban:1998eg}
  M.~Urban , M.~Buballa, R.~Rapp and J.~Wambach,
 Nucl.\ Phys.\ A {\bf 641}, 433 (1998).

\bibitem{Rapp:2000pe}
  R.~Rapp,
  Phys.\ Rev.\ C {\bf 63}, 054907 (2001).

\bibitem{Dey:1990ba}
  M.~Dey, V.~L.~Eletsky and B.~L.~Ioffe,
  Phys.\ Lett.\ B {\bf 252}, 620 (1990).

\bibitem{SYZ96}
J.V.~Steele, H.~Yamagishi, and I.~Zahed, Phys. Lett. B \textbf{384},
255 (1996).

\bibitem{Ding:2010ga}
  H.-T.~Ding {\it et al.}, 
  Phys.\ Rev.\ D {\bf 83}, 034504 (2011).

\bibitem{Brandt:2012jc} 
  B.B.~Brandt, A.~Francis, H.B.~Meyer and H.~Wittig,
  JHEP {\bf 1303}, 100 (2013).

\bibitem{Kapusta:1991qp}
  J.I.~Kapusta, P.~Lichard and D.~Seibert,
  Phys.\ Rev.\ D {\bf 44}, 2774 (1991)
  [Erratum-ibid.\ D {\bf 47}, 4171 (1993)].

\bibitem{Braaten:1990wp}
  E.~Braaten, R.D.~Pisarski and T.-C.~Yuan,
  Phys.\ Rev.\ Lett.\  {\bf 64}, 2242 (1990).

\bibitem{vanHees:2011vb}
  H.~van Hees, C.~Gale and R.~Rapp,
  Phys.\ Rev.\ C {\bf 84}, 054906 (2011).

\bibitem{vanHees:2014ida} 
  H.~van Hees, M.~He and R.~Rapp,
  arXiv:1404.2846 [nucl-th].

\bibitem{Kumar:2014tca}
  L.~Kumar [STAR Collaboration],
  arXiv:1408.4209[nucl-ex].

\bibitem{He:2011zx}
  M.~He, R.J.~Fries and R.~Rapp,
  Phys.\ Rev.\ C {\bf 85}, 044911 (2012).

\bibitem{Borsanyi:2010cj}
S.~Borsanyi {\it et al.}, 
  JHEP {\bf 1011}, 077 (2010).

\bibitem{Cheng:2009zi}
M.~Cheng {\it et al.}, 
  Phys.\ Rev.\ D {\bf 81}, 054504 (2010).

\bibitem{Endrodi:2011gv} 
  G.~Endrodi, Z.~Fodor, S.~D.~Katz and K.~K.~Szabo,
  JHEP {\bf 1104}, 001 (2011).

\bibitem{Kaczmarek:2011zz} 
  O.~Kaczmarek {\it et al.}, 
  Phys.\ Rev.\ D {\bf 83}, 014504 (2011).

\bibitem{Borsanyi:2012cr} 
S.~Borsanyi {\it et al.}, 
JHEP {\bf 1208}, 053 (2012).

\bibitem{Hegde:2014sta} 
  P.~Hegde {\it et al.},  
  arXiv:1408.6305 [hep-lat].

\bibitem{BraunMunzinger:2011ze} 
  P.~Braun-Munzinger and J.~Stachel,
  arXiv:1101.3167 [nucl-th].

\bibitem{Arnaldi:2008fw}
  R.~Arnaldi {\it et al.}  [NA60 Collaboration],
  Eur. Phys. J. C {\bf 61}, 711 (2009).

\bibitem{Agakichiev:2005ai}
  G.~Agakichiev {\it et al.}  [CERES Collaboration],
  Eur.\ Phys.\ J.\ C {\bf 41}, 475 (2005).

\bibitem{Renk:2006qr} 
  T.~Renk and J.~Ruppert,
  Phys.\ Rev.\ C {\bf 77}, 024907 (2008).

\bibitem{Dusling:2006yv} 
  K.~Dusling, D.~Teaney and I.~Zahed,
  Phys.\ Rev.\ C {\bf 75}, 024908 (2007).

\bibitem{Alam:2009da} 
  J.~K.~Nayak, J.~e.~Alam, T.~Hirano, S.~Sarkar and B.~Sinha,
  Phys.\ Rev.\ C {\bf 85}, 064906 (2012).

\bibitem{Linnyk:2011hz} 
  O.~Linnyk, E.~L.~Bratkovskaya, V.~Ozvenchuk, W.~Cassing and C.~M.~Ko,
  Phys.\ Rev.\ C {\bf 84}, 054917 (2011).

\bibitem{Rapp:2013ema} 
  R.~Rapp,
  PoS CPOD {\bf 2013}, 008 (2013).

\end{thebibliography}
\end{document}